\documentclass[prc,aps,floatfix,nofootinbib,twocolumn,superscriptaddress]{revtex4} 

\usepackage{graphicx}
\usepackage{color}
\usepackage{enumerate}
\usepackage{amssymb}
\usepackage{amsmath}
\usepackage{amsfonts}
\usepackage{bbm}
\usepackage{dsfont}
\def\be{\begin{equation}} 
\def\ee{\end{equation}} 

\newcommand{\vecbold}[1]{\mbox{\boldmath $#1$}}

\begin{document}

\title{Role of momentum in the generator-coordinate method applied
to barrier penetration}

\author{K. Hagino}
\affiliation{ 
Department of Physics, Kyoto University, Kyoto 606-8502,  Japan} 

\author{G.F. Bertsch}
\affiliation{ 
Department of Physics and Institute for Nuclear Theory, Box 351560, 
University of Washington, Seattle, Washington 98915, USA}

\begin{abstract}
Nuclear fission at barrier-top energies is conventionally modeled by
a one-dimensional Schr\"odinger equation applied to internal fission 
channels,  but that treatment
is hard to justify in the  configuration-interaction
approach to nuclear Hamiltonians.
Here we show that inclusion of states of finite momentum by the
Generator Coordinate Method (GCM) considerably 
extends the range of energies at which 
GCM-based Hamiltonians could reproduce the Schr\"odinger treatment.
The transmission probabilities for crossing the barrier are calculated
by a discrete version of Kohn's variational method, which may also be
useful for other systems of interacting fermions.
\end{abstract}

\maketitle

\section{Introduction} 

As discussed in our previous papers in this series\cite{kh23,gb24}, there are 
many unanswered questions about the dynamics of nuclear
fission that require a fully microscopic theory to answer definitively.  Most
fundamentally, should one treat the dynamics as largely diffusive as in
Kramer's model \cite{kr40}, or as motion along a collective coordinate
as in Bohr and Wheeler's original transition-state model \cite{BW}?     
Given the success of the
configuration-interaction (CI) framework for calculating energies and
spectroscopic properties, that approach should be  
explored for fission theory as well.
Our previous papers have advocated using the CI framework
and particularly the Generator Coordinate Method (GCM) to
build a reaction theory of fission that might possibly address the
questions.  The generator-coordinate method (GCM) has been a very useful tool
for constructing many-particle wave functions, particularly as applied to deformed systems undergoing
shape changes; see  Ref. \cite{bo90} for an early study.

Absent a microscopic theory,  barrier traversal has been treated by 
one-dimensional Schr\"odinger dynamics since the  
pioneering work of Hill and Wheeler \cite{HW}, and continuing up
to the present era \cite{si16,ko23,be23,kawano2023}.
Since that approach is so well established, one can ask what is 
required for the discrete-basis CI Hamiltonian to reproduce the 
Schr\"odinger results.
In our 
previous work,  we showed how a Hamiltonian constructed from a chain of
GCM states could produce the transmission probabilities in a limited
range of energies,  requiring barrier heights no higher than a few
(2-3) times larger than the zero-point energies of the states.  
Here we show that adding a second state on each site of the chain
can double the useful energy range.  That is more than enough to 
reproduce the transmission probabilities calculated
in the Schr\"odinger approach. 

This article is organized as follows. We first summarize
the construction of a simplified  Hamiltonian compatible with
the GCM framework.
Next we extend the space to include 
higher momentum states and assess their fidelity as momentum
eigenstates.  Then we formulate a computationally tractable reaction 
theory following Kohn's \cite{ko48} employment of asymptotic channel
wave functions.
Transmission probabilities calculated in that scheme are
reported in Sec. \ref{IV} and compared with the Schr\"odinger
results.

\section{The discrete-basis Hamiltonian}

  The target Schr\"odinger Hamiltonian  has the usual
form
\be
    H = -\frac{\hbar^2}{2 M}\frac{\partial^2}{\partial x^2} + V(x),
\label{Hx}
\ee 
where $x$ is a collective coordinate defined by the GCM constraints.
The $M$ in the kinetic term is the collective inertia associated with
the coordinate and $V(x)$ is the barrier potential.
To mimic a discrete-basis GCM Hamiltonian  
basis,  we assume that the wave function can be 
be factorized into a Gaussian wave packet along the collective
coordinate times an internal wave function that need not be
specified.  The space consists of a mesh of these wave packets
centered on equally spaced points $\Delta x$ along the collective coordinate.
The model is completely determined by the Hamiltonian and overlap
matrix elements between these states,  $H_{ij} = \langle i | H | j \rangle$
and $N_{ij} = \langle i | j \rangle$.  We assume these elements 
depend only on the separation between the two states $i$ and $j$ along the
chain, apart from the barrier potential $V$.  The Gaussian wave packet
is parameterized as the ground-state harmonic oscillator wave function
\be
\phi_{0,x_i}(x) = \frac{1}{\pi^{1/4} s^{1/2}}e^{- (x-x_i)^2/2 s^2}.
\label{phi-def}
\ee
Here the subscript $x_i$ denotes the center point of the collective 
coordinate in the GCM wave packet.  With the above definition the
overlap matrix elements are 
\be
\label{Nij}
N_{i,j} = e^{-(i-j)^2\Delta x^2/4 s^2}.
\ee
The Hamiltonian matrix elements are parameterized in the well-known
Gaussian overlap approximation (GOA) as \cite{goa}
\be
H_{i,j} = N_{i,j} E_q ( 1 - (i-j)^2\Delta x^2 /s^2).
\label{Hij}
\ee
Here 
\be 
E_q = \frac{\hbar^2}{4\,M\,s^2}
\label{eq:zpe}
\ee
is the zero-point kinetic energy in the collective coordinate.
Of course in an actual GCM
Hamiltonian the matrix elements will not be so regular as assumed in
Eqs. (\ref{Nij},\ref{Hij}).

Note that the parameter $\Delta x$ is purely numerical and must be
chosen with some care.  If it is too small the diagonalization
of the matrix Hamiltonian equation
\be
\vecbold{H} \psi = E \vecbold{N}\psi
\label{eq:HNE}
\ee
becomes numerically unstable. 
On the other hand, if it is too
large the space of configurations along the collective path is not
well sampled.  We choose the parameter
value  $\Delta x = 5^{1/2}s $ as a somewhat arbitrary compromise between
these considerations \cite{BY}.

The new ingredient in this work is the explicit introduction of 
momentum in the wave function at each site.  This can be done in different
ways, depending on how the configurations are constructed \cite{kr17,
efros19}. In
the GCM framework one could add a momentum constraint or simply take
derivatives with respect to $x_i$, the central position of the wave packet
at site $i$.  The resulting wave packet is just the first excited
harmonic oscillator wave function,
\be
\phi_{1,x_i}(x) = 2^{1/2}(x - x_i) e^{-(x-x_i)^2/2 s^2}/\pi^{1/4}.
\ee
Our space of states will be defined with $\phi_{1,x_1}$ as the
second state on the site. 

\section{Construction of plane waves and their properties}
Wave functions simulating plane waves are trivial to construct for states
with a fixed internal structure
on equally spaced grid points;  the amplitudes on adjacent sites
are related by identical phase factors $e^{i \theta}$. It
is less obvious how to build them when there are two or 
more states at each grid point.  In this work we determine
them from eigensolutions of the kinetic Hamiltonian as follows. In
the first step, a periodic matrix Hamiltonian
is defined on a ring  of $N_q$ mesh points.  Translational symmetry is
imposed by requiring $H$ and $N$ to depend only on the difference
between the site variables, that is
\be
\langle \mu,x_i|H| \mu',x_{j}\rangle = H_{|i-j|,\mu,\mu'}
\ee
Here $\mu$ labels the states on a site. The dimension of the
space is thus $N_q N_s$ where $N_s$ is the number of states
on each site.
The overlap matrix is constructed in the same way. The matrices
are band-diagonal with nonzero matrix elements extending to 
$N_{\rm od}$ sites on either side of the diagonal.  For our
numerical study we include next-to-nearest neighbor interactions,
$N_{\rm od} = 2$.  Details of the
matrix elements are given in the Appendix A.  
  
Next one solves the generalized eigenvalue equation Eq. (\ref{eq:HNE})
expressing the eigenstates in terms of amplitudes
$\vec{f_0}$ and $\vec{f_1}$: 
\be
\psi = \sum_i \left(f_{0,i} \phi_{0,x_i} + f_{1,i} \phi_{1,x_i}\right).
\ee 
All the eigenvalues and amplitudes $f$ are real, and all but two of them
have a partner with the same eigenvalue.  In fact, these eigenvectors and
energies can be calculated much more simply from the formulas derived in
Appendix B.

\begin{figure}[htb] 
\begin{center} 
\includegraphics[width=0.9\columnwidth]{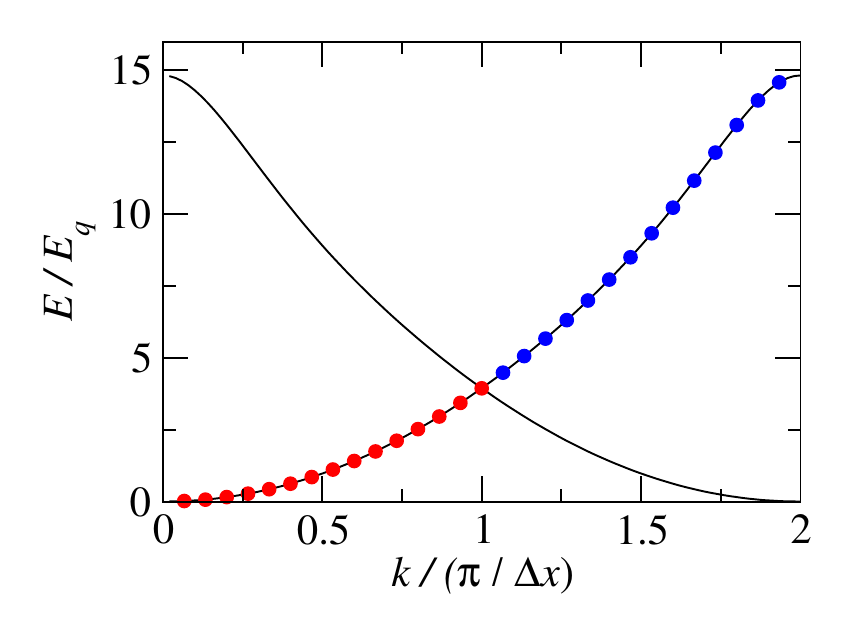} 
\caption{
Energies  of plane-wave states in the discrete-basis formalism, as
calculated by diagonalizing the ring Hamiltonian with $N_q =30$ sites.
The Hamiltonian and overlap 
matrices include elements up to second nearest neighbor interactions,
$N_{\rm od} = 2$.
Red dots and blue dots show the energies of the lower half and upper
half of the eigenenergy spectrum.  See text for the assignment of $k$
values to the eigenstates.  The lines show the continuous dispersion
curve Eqs. (\ref{eq-app:HW2},B7-8) derived in Appendix B.    See
Fig. \ref{Evsk2} for comparison with the Schr\"odinger spectrum.
\label{Evsk1}} 
\end{center} 
\end{figure}

The traveling waves are constructed from
paired eigenvectors $\psi_1,\psi_2$ as 
\begin{eqnarray}
    \psi_k &=& \psi_1  + i \psi_2 \\
    \psi_{-k} &=& \psi_1  - i \psi_2.
\end{eqnarray}
The subscript $\pm k$ denotes the momentum of the state,  which is still to
be determined.
Within each of these traveling waves
the amplitudes at adjacent sites are related by
$e^{\pm i \theta}$.  Due to the ring structure of the Hamiltonian the phase $\theta$ 
is quantized with the form   
\be
\theta = \pm  m \frac{\pi}{s N_q}.
\ee
where the  $m$ can be 
restricted to the range $-N_q < m < N_q$. For Hamiltonians with a single state
on site,  the states approximate eigenstates of the momentum operator with
$k \approx  \theta/\Delta x$. In the Hamiltonian  with an added momentum state
on the sites the number of eigenstates
is doubled.  The lower energy states become more accurate approximations to
the same momenta in the range $0 < |k| < \pi/\Delta x$.  However, 
the phase change for the higher energy states  is better interpreted 
by the assignment $k \approx (\theta +\pi) \Delta x$.  The situation is depicted in 
Fig. \ref{Evsk1}.  The momentum  assignments based solely on the periodicity from
site to site are shown as the red dots in the figure.  The blue dots
in the upper range of energies have the same site-to-site periodicity.
However, 
due to the on-site structure the momenta of these wave functions follow more 
closely the continuation of the low-energy dispersion curve.  
The continuous curve is drawn using the method presented in Appendix B.
We have verified the momentum assignments by 
evaluating the expectation value of the momentum operator
$\partial/i\partial x$.  The agreement is approximate but close enough
for our purposes.

The amplitudes of the eigenstates on each site can be
expressed as
\be
(f_{0,m},f_{1,m}) = e^{i m \pi/N_q} (1,\alpha)
\ee
times an overall factor. Here $m$ labels the site and 
$\alpha$ controls the  mixing of the two states
on the site. The mixing amplitude is independent of the site and is purely imaginary
for our Hamiltonian.   The assigned momentum of this eigenstate is 
$k = m \pi /(N_q\Delta x)$ or $m\pi /(N_q\Delta x) \pm\pi/N_q$ depending on whether 
the energy is in the lower or upper half of the spectrum.
Note also that right-moving and left-moving plane waves $\psi_k$
and $\psi_{-k}$ are related by complex conjugation.  

Fig. \ref{Evsk1}  shows the dispersion curve 
for the ring Hamiltonian with next-to-nearest neighbor interactions
and two states on each site.
In this calculation,
the mesh spacing was set to $\Delta x = 5^{1/2} s$ as in the
previous work.  Besides diagonal elements, the matrices include 
off-diagonal elements up to second nearest neighbors.  The figure
shows that the two-state construction considerably extends the
range of momenta that are covered. 

\begin{figure}[htb] 
\begin{center} 
\includegraphics[trim= 0   0  0 0,
clip=true,width=0.8\columnwidth]{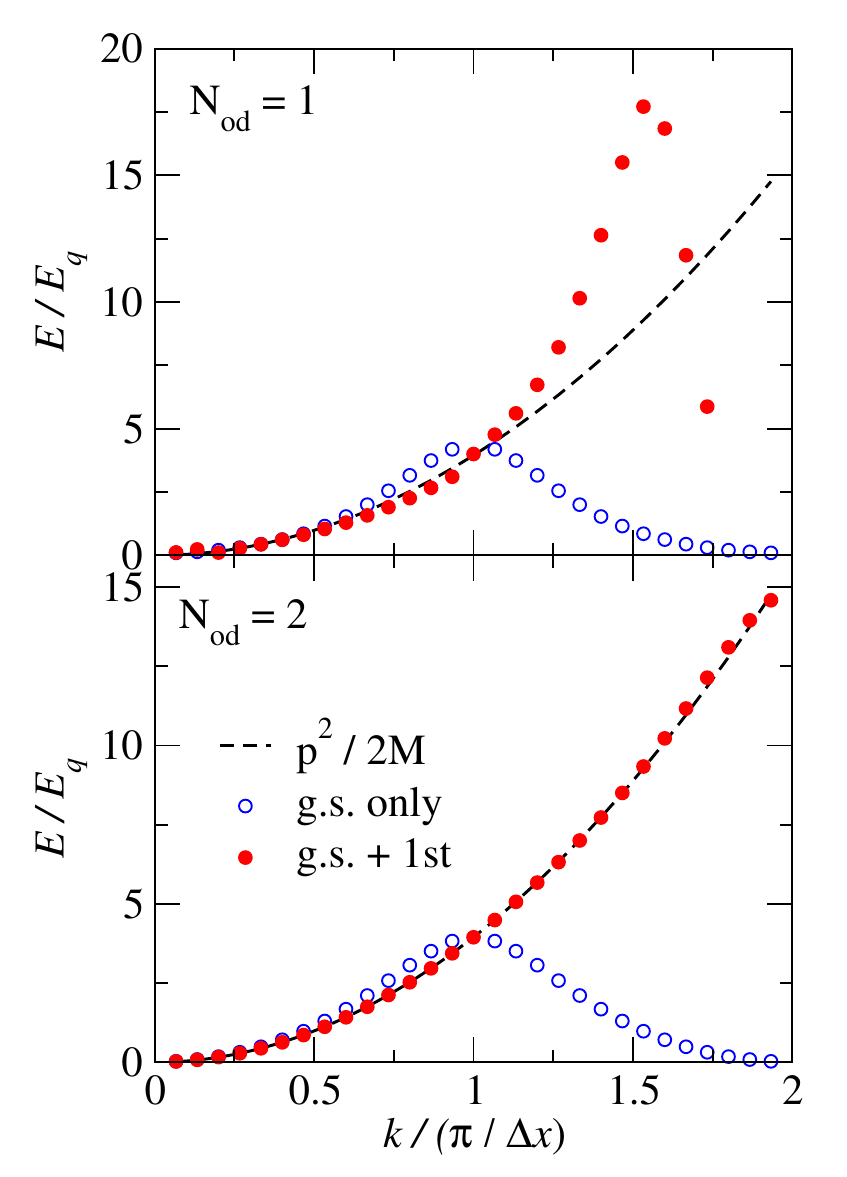} 
\caption{
Energies  of plane-wave states in the discrete-basis formalism 
in the $N_s = 1$ and $2$ approximations.  The upper and the lower 
panels show the results with $N_{\rm od}=1$ and 2, respectively.
The filled
red circles show the energies with ther Hamiltonian parameters
the same as in Fig. 1.  Open circles show the energies with
$N_s= 1$.  The dashed line shows the Schr\"odinger dispersion curve, 
Eq. (\ref{Ek2}).}
\label{Evsk2}
\end{center} 
\end{figure}

Fig. \ref{Evsk2} compares the present Hamiltonian to the Schr\"odinger
dispersion curve
\be
E = \frac{1}{2 M} k^2.
\label{Ek2} 
\ee
The agreement is quite close, and only deviates significantly at the
highest momenta.
Also shown in the figure are the dispersion curves for the
$N_{\rm od} =1$ Hamiltonian studied in Ref. \cite{gb24}. Adding the
second state on a site significantly improves the agreement
for $ |k| < \pi/\Delta x$ as well extending the domain
to $ k\ < 2 \pi/\Delta x$.

\section{Barrier penetration in the Kohn reaction theory}
\label{IV}
\subsection{Formalism}
		
The barrier penetration problem may be treated in reaction theory as 
a system with two active exit channels, namely the transmitted channel
on the other side of the barrier and the reflected flux within the 
entrance channel.  There are several reaction formalisms based on the
$S$-matrix to treat discrete-basis Hamiltonians.
If Hamiltonian in the channel space has the simple form of the
one-dimensional nearest-neighbor hopping  model,
the theory is
quite straightforward and it provides numerically exact $S$-matrix algorithms
to calculate reaction observables.

This is not the case when the channel Hamiltonian has a more complex
structure.  With more than one state on a site,
the detailed structure
of the channel wave functions has to be included in the boundary conditions
on the Hamiltonian.  In principle this can be carried out in the
$S$-matrix or the $K$-matrix formalism, but to deliver numerically
exact observables the formalisms require principal-value integrals
over states in the interaction regime and channel wave functions. None
of our attempts to solve the barrier problem with the wave functions
described in our earlier publications were successful.  However, we did succeed in getting
reasonable results based on Kohn's variational principle \cite{ko48,hu48}.
In fact this method has been previously applied to nucleus-nucleus 
collisions with Hamiltonians constructed by the GCM \cite{be75,ka77,vi20} 
as well as the  no-core shell model \cite{efros19,sharaf2024}. 
The procedure is described below.

The essence of the method is to include at the outset
the channel wave functions in the asymptotic region, taking
their amplitudes as specific unknown 
variables to be calculated.
In principle it does not matter that the asymptotic
channel wave function is wrong in the interaction region because
there are other variables in the wave function that will correct them.

In the present problem, there are three channel wave functions to 
be included, namely:   the incoming plane wave $\psi^L_k$, the reflected plane wave
in the same space $\psi^L_{-k}$, and the outgoing transmitted wave
$\psi^R_k$. There
is no incoming wave in the transmitted channel, so it does not
appear in the formalism.  Each state on a mesh point of the collective
coordinate belongs to one of three regions:  asymptotic on the left hand
side, asymptotic on the right hand side, or in the interacting 
region in between.  The fundamental matrix 
equation to be solved is Eq. (\ref{eq:HNE}) which we now write
in the form
\be 
 \vecbold{H}'\psi = (\vecbold{H}-E\vecbold{N})\psi = 0.
\label{Hp2}
\ee
The goal to solve the equation is approached row
by row in  the matrix-vector  product $\vecbold{H}'\psi$.
The full solution will have the form
\be
\psi = \psi^L_k + c_1 \psi^L_{-k} + c_2 \psi^R_k + 
\sum_{n = 1}^{N_\phi/2} \sum_{\mu = 0,1} f_{\mu,n}  \phi_{\mu,n}.
\label{solution}
\ee 
Similar equations to Eqs. (\ref{Hp2},\ref{solution}) appear in various forms
in many publications using the variational approach to reaction theory.
The undetermined amplitudes $c_1$ and $c_2$ in Eq.(\ref{solution}) are associated with the channels
and the $N_\phi$ amplitudes $f_{\mu,n}$ are associated with individual GCM states
in the interaction region.  

\begin{figure}[htb] 
\begin{center} 
\includegraphics[width=1\columnwidth]{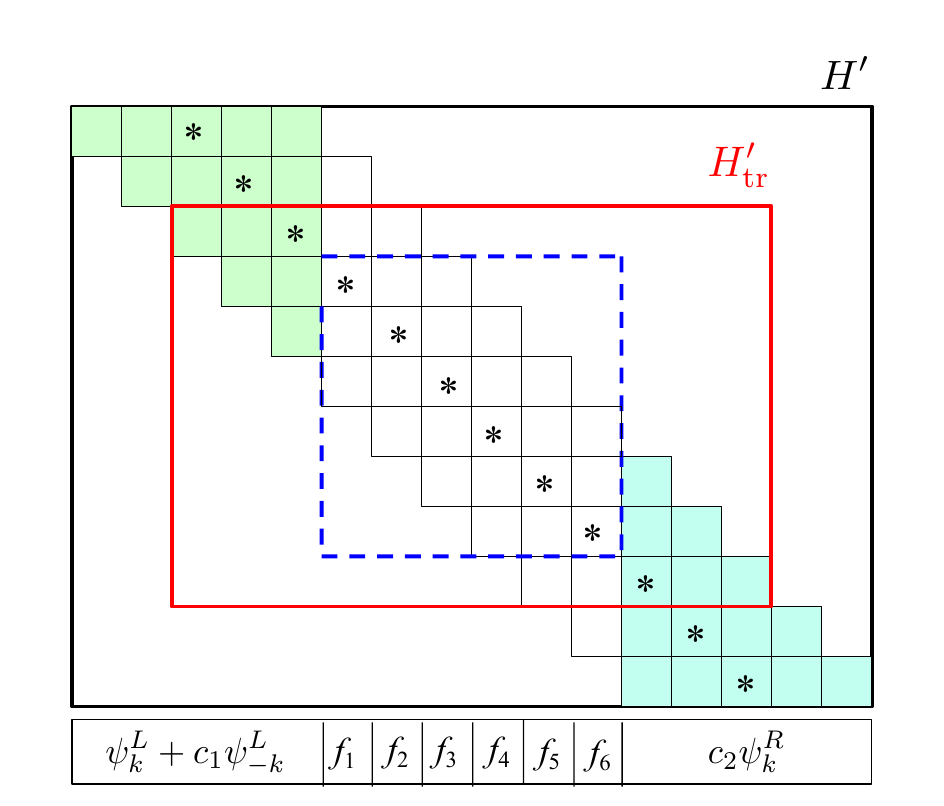} 
\caption{Active elements of the Hamiltonian for solving the
Eq. (\ref{Hp2}) having  parameters $(N_s,N_{\rm od}) = (1,1)$.
The truncated matrix $\vecbold{H}'_{\rm tr}$ is 
outlined in red, and the interior elements are encased by the blue
dotted square.  Active elements in the kinetic part of $\vecbold{H}'$ are indicated
by the small  boxes. See text for additional details.  explanation.
\label{Hpchart}} 
\end{center} 
\end{figure}

To see how Eq. (\ref{Hp2}) is solved, consider a simplified 
one having $N_{\rm od} = 2$ and only 
a single state on each site.
The active elements of the Hamiltonian $\vecbold{H}'$ are depicted in Fig. \ref{Hpchart}.
The top row of the Hamiltonian is zero except for the 5 entries
for the kinetic energy operator.  It
acts on the left hand channel wave function constructed as 
a linear combination of $\psi^L_k$ and $\psi^L_{-k}$.  Since
both wave functions are asymptotic solutions, the row
condition $H'\psi =0$ is already satisfied.
For the next row down, the Hamiltonian
may have some contribution from the interaction potential since
the kinetic operator extends into the interaction region.
Similarly, the next row below extends to two states in that
region.    The situation is the same on the right hand
side of the interaction zone.  The lowest row depicted is
completely in the asymptotic region, and the Eq. (6) is
automatically satisfied by $\psi^R_k$.  The Hamiltonian $\vecbold{H}'$ depicted here also has 
$N_\phi = 6$ internal states giving the same number of amplitudes $f_1...f_6$
to be determined. Besides those amplitudes 
for the states in the interaction region, we need to determine
the amplitudes $c_1$ and $c_2$ for the outgoing channel wave functions.
This gives a total $N_\phi +2 = 8$ unknowns.  However there are 10
equations for the 10 rows active in the interaction region,
so it appears that the problem is over-determined.

That cannot be a fatal flaw because we know that a solution
exists.  So there must be a hidden relationship between the
amplitudes that allows all 10 row equations to be satisfied.
We can find the amplitudes by ignoring two of those rows
and solve the algebraic problem of solving 8 equations for
8 linear unknowns.  In our implementation, we drop the rows immediately adjacent
to the asymptotic region to set up a $(N_\phi +2)\times (N_\phi +2)$
matrix to be inverted.  In detail, the matrix $\vecbold{H}'$ is truncated to 
$\vecbold{H}'_{tr}$ with $N_\phi+2$ 
rows and $N_\phi +6$ columns.  Then the unknown amplitudes in
Eq. (\ref{solution}) are obtained by the matrix inversion of the
matrix-vector equation
\begin{eqnarray}
&&
(\vecbold{H}'_{\rm tr}.\psi^L_{-k},~\vecbold{H}'_{\rm tr}.\phi_{1},\cdots,\vecbold{H}'_{\rm tr}.\phi_{N_\phi},
~\vecbold{H}'_{\rm tr}.\psi^R_{k}
)
\left(
\begin{array}{c}
c_1 \\
f_1 \\
\vdots \\
f_{N_\phi} \\
c_2
\end{array}
\right) \nonumber \\
&& = - \vecbold{H}'_{tr}.\psi^L_k
\label{Htr-cf}
\end{eqnarray}  
Since the solution of the physical problem is unique, it must
be the case that the resulting wave function Eq. (\ref{solution})
also satisfies the condition Eq. (\ref{Hp2}) for the rows that
were dropped. We have checked that that is indeed the 
case for the numerical examples shown below.  
Having solved Eq. (\ref{Htr-cf}) for the unknown amplitudes,
the transmission and reflection probabilities $T$  and $R$ are
obtained from the ratio of channel normalizations $c_1,c_2$ to the
incoming wave normalization (equal to one in Eq. (\ref{solution})),
\be
R=|c_1|^2,~~~
T = |c_2|^2.  
\ee
Since there is no absorption of flux, the relation 
$T+R=1$ should hold. 

\subsection{An example}
\label{IVb}
To recapitulate the essential ingredients of the present model,
we assume that the GCM states satisfy the Gaussian 
overlap approximation and construct the $\vecbold{H}$ and $\vecbold{N}$ matrices accordingly.
The two parameters controlling the matrix elements of $\vecbold{N}$ are the
width of the Gaussian,  $s$ in Eq. (\ref{phi-def}) 
and the mesh spacing $\Delta x$ on the collective coordinate.  As in our previous studies,
distances along the collective coordinate are expressed in units of
$s$ and the mesh spacing is taken as $ 5^{1/2}$ in those units.
The kinetic energy operator is the usual one with the unit of
energy taken as the zero-point energy of the GCM ground state, 
Eq. (\ref{eq:zpe}).
The barrier potential is parameterized as
\be
V(x)  = V_0 e^{- x^2/ 2 \sigma^2}
\ee
with $V_0$ the height of the barrier and $\sigma$ its width. 
For this example the width parameter 
is set to $\sigma = 2.0$.  
All the needed
matrix elements are Gaussian integrals; the formulas are given
in the Appendix A.  The potential is effective over a region extending
to about 10 sites on the chain of GCM states; we take 
$N_q = 30$ sites and  $N_\phi =60$ 
basis states in and around the barrier to define the interior
region. To define  $\vecbold{H}'_{\rm tr}$ the full Hamiltonian is truncated to
$N_r = N_\phi +2$ rows, equal to the number of unknown amplitudes. The
number of columns must be  $N_\phi +6$ or larger to accommodate 
the nonzero Hamiltonian matrix elements in the top and bottom rows.
The minimum dimensions of $\vecbold{H}'_{\rm tr}$ are thus $62\times66$.

\begin{figure}[htb] 
\begin{center} 
\includegraphics[width=0.9\columnwidth]{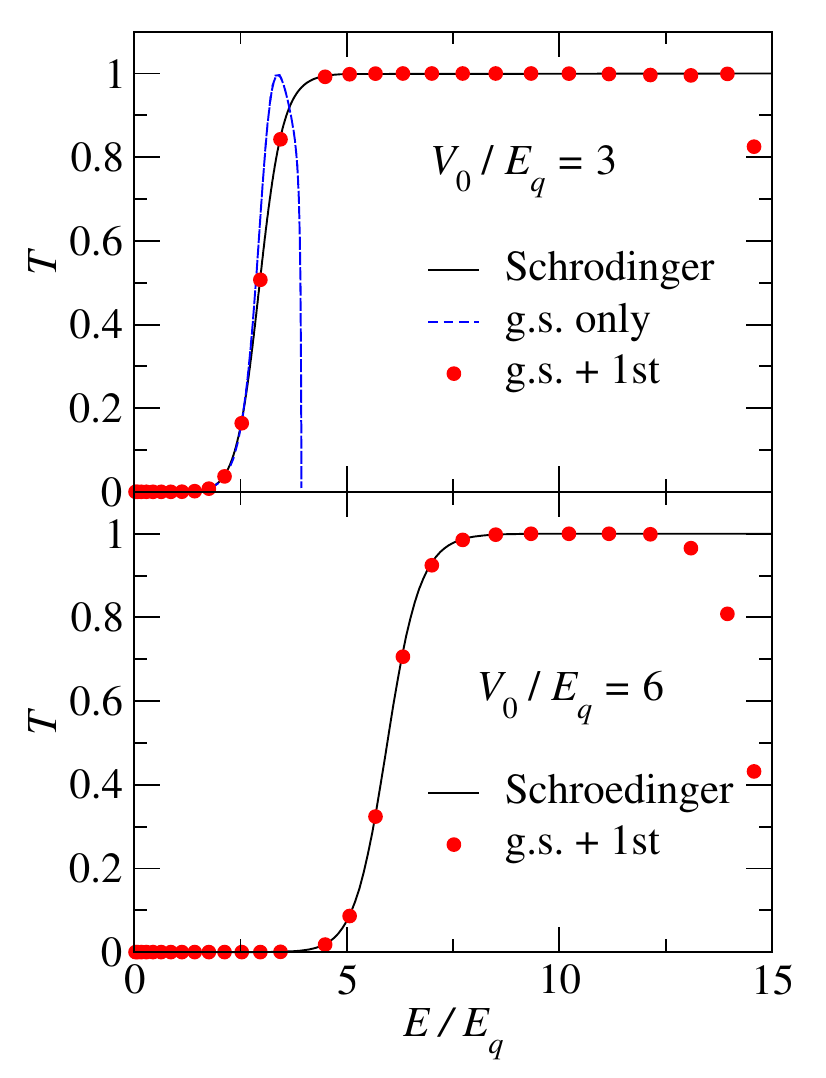} 
\caption{
Transmission probabilities for traversing a barrier computed
in the discrete-basis Hamiltonian approximation to the one-dimensional
continuum Hamiltonian Eq. (\ref{Hx}).  The results of the full treatment
with $N_s=2$ and $N_{\rm od} =2$ are shown by the red circles.  The results 
from direct integration of the Schr\"odinger equation are shown by
the solid line.  The upper
panel with barrier height $V_0 = 3 E_q$ shows a comparison with
the previous published calculation which had only the $\phi_0$ states
in the basis. In the lower panel the barrier height is $V_0 = 6 E_q$,
which is beyond the range of the $N_s = 1$ treatment.}
\label{kohn36}
\end{center} 
\end{figure}

Fig. \ref{kohn36} shows the calculated transmission probability for
the above set of Hamiltonian parameters, with a comparison to the
direct integration of the Schr\"odinger equation Eq. (\ref{Hx}) and
to a previous calculation in the discrete-basis formalism.
One sees that present theory is much improved for
energies $E/E_q < 3$ and that the catastrophic failure of the previous
treatment above that energy is prevented.  
In the lower panel one 
sees that the present treatment gives excellent agreement up to at
least $E/E_q \approx 10$. 
That would provide an acceptable energy range
to study the role of fission channels in 
actinide nuclei.

One needs to choose $N_\phi+2$ rows $i$ from the original 
Hamiltonian matrix $H'_{ij}$ to define the truncated Hamiltonian, $\vecbold{H}'_{\rm tr}$. 
These rows can be arbitrary as long as 
there is a coupling between the asymptotic wave 
functions and the internal wave functions. 
The most appropriate choice would be 
to take $N_\phi$ internal rows with 2 extra asymptotic rows on the both sides of a barrier.   
When there is only one state at each site, this provides a unique choice for $\vecbold{H}'_{\rm tr}$. 
On the other hand, when there are two states at each site, there is an ambiguity, 
as there can be four possible choices, that is, $(\phi_{L,0},\phi_{R,0})$, $(\phi_{L,0},\phi_{R,1})$, 
$(\phi_{L,1},\phi_{R,0})$, and $(\phi_{L,1},\phi_{R,1})$ with an obvious notation. 
We have confirmed that those four choices lead to the same result when $N_\phi$ is large, such as 
$N_\phi=200$. In contrast, when $N_\phi$ is small, such as $N_\phi=20$, only the choice 
$(\phi_{L,0},\phi_{R,0})$ leads to numerically stable solutions. 
We found that for the other choices, there is one or two small eigenvalues 
of the matrix that has to be inverted in Eq. (\ref{Htr-cf}). 
We anticipate that this is the origin of the numerical instability, since all the 
eigenvalues are large for the choice of $(\phi_{L,0},\phi_{R,0})$. 

\section{Discussion}
As a study of formalisms for reaction theory, our results extend the range of 
phenomena that can be studied within the CI approach using the GCM to 
construct the configurations.  Kohn's method gives a general technique
for carrying out numerically accurate calculations of 
discrete-basis Hamiltonians from a  GCM construction. 
We showed how the formalism can provide reaction branching ratios
when there is another exit channel besides the entrance channel.

For nuclear physics and the question about the role of internal channels
in fission, the crucial parameters are barrier height and the zero-point energies of the
configurations.  It appears that these parameters
are in a safe region for applying the Kohn formalism and with the
present numerical parameters.  

However, it should be mentioned that algorithms based on Kohn's treatment
are subject to numerical instabilities \cite{dr21}.  This is true for our
discrete-basis approach as well.  The set of equations Eq.
(\ref{Hp2},\ref{solution})
is overdetermined, requiring us to ignore some rows in setting up the matrix 
$\vecbold{H}'_{\rm tr}$ to be inverted.  With a poor choice the matrix is nearly singular and the
calculated wave function is not reliable\footnote{In Ref. \cite{od24},
various combinations of rows are tested to maximize the determinant of the
matrix.}.  One test for a reliable result
is to check flux conservation, i.e.,  the transmission and reflection 
probability satisfy $T + R = 1$.  We have also found that the method is
less sensitive to the choice of rows if the nominal
interior space is extended into the channel space.
 
The computer codes to calculate the $N_s=2$ data plotted in Fig. 2
and 4 are provided in the Supplementary Material. 
Also included is a code to compute the 
plane-wave energy $E$ from Eqs. (B4), (B7), and (B8). 

\begin{acknowledgments}
This work was supported in part by
JSPS KAKENHI Grant Number JP23K03414.
\end{acknowledgments}

\appendix

\section{Matrix elements of $H,N$ and $V$}

The Gaussian integrals that arise in calculating the elements of the
discrete-basis matrices are evaluated in this appendix. The two
physical parameters are $s$ governing the size of the wave packet
and $E_q$, the zero-point energy of the wave packet. The relation
to the inertial mass $M$ along the collective coordinate is
$E_q = \hbar^2/4 M_q s^2 $.  In the formulas
below, lengths are expressed in units of $s$.

For the overlap matrix $N$, we write its elements as
$\langle \phi_n(x_2)| \phi_{n'}(x_1) \rangle  = N_{n,n'}(\Delta x)$ 
where $\Delta x  = x_1 - x_2$.  The matrix elements are:
\be 
N_{00}(\Delta x) = e^{-\Delta x^2/4 }
\ee
\be
N_{01}(\Delta x) = -N_{10}(\Delta x) = - 2^{-1/2} \Delta x N_{00}(\Delta
x)
\ee
\be
N_{11}(\Delta x) = (1-(\Delta x)^2/2 ) N_{00}(\Delta x)
\ee

For the kinetic term $T$ in the Hamiltonian, the matrix matrix elements are

\be
T_{00}(\Delta x) = E_q (1 - (\Delta x)^2/2) N_{00}(\Delta x)
\ee
\be
T_{01}(\Delta x) = E_q \left(3-(\Delta x)^2/2\right) N_{01}(\Delta x)
\ee
\be
T_{11}(\Delta x) = E_q\left(3  -\frac{3}{2}(\Delta x)^2  
\frac{1-(\Delta x)^2/6}{1-(\Delta x)^2/2}\right)
N_{11}(\Delta x)
\ee

Finally the matrix elements of the barrier potential $V(x) = V_0 e^{-x^2/2\sigma^2}$:
are:
\be
V_{00}(x_1,x_2) = V_0\sqrt{\frac{2 \sigma^2}{2 \sigma^2 +1}} A
\label{V00}
\ee
\be
V_{01}(x_1,x_2) = 2V_0\frac{-\sigma^2 \Delta x +x_2}{(2 +1/\sigma^2)^{1/2} (1 + 2
\sigma^2)} A   
\label{V01}
\ee
\begin{eqnarray}
V_{11}(x_1,x_2) &=& 2^{3/2} \sigma V_0 \nonumber \\
&&\times\frac{\sigma^4(2-(\Delta x)^2)+
\sigma^2 (1-(\Delta x)^2)+x_1 x_2}
{(1+2 \sigma^2)^{5/2}} A \nonumber \\
\label{V11}
\end{eqnarray}
where $A$ is defined
\be
A = \exp\left(-\frac{(1+\sigma^2)({x_1}^2+{x_2}^2) - 2 \sigma^2 x_1 x_2}
{2+4\sigma^2}\right). 
\ee
Eqs. (\ref{V00}) and (\ref{V11}) can be conveniently written in factorizable form with
the corresponding overlap $N_{n,n'}$ as
\be
V_{00}(x_1,x_2) = \tilde A N_{00} (\Delta x)
\ee
and 
\begin{eqnarray}
V_{11}(x_1,x_2) &=& \left(1 + \frac{1}{1-(\Delta x)^2/2}\left(
\frac{(x_1+x_2)^2}{2(1+2 \sigma^2)^2} - \frac{1}{1+2 \sigma^2}\right) 
\right) \nonumber \\
&&\times \tilde A N_{11}(\Delta x),
\end{eqnarray}
where
\be
\tilde A = V_0\sqrt{\frac{2 \sigma^2}{1 + 2 \sigma^2}} e^{-\frac{1}{4}\frac{(x_1+x_2)^2}
{1 + 2 \sigma^2}}.
\ee
The factorizable form for Eq. (\ref{V01}) is indeterminate at $\Delta x = 0$
because $N_{01}(0) = 0$.

\section{Structure of plane-wave states by the GCM}

In this appendix we derive formulas for the energies and structure
of GCM continuum wave functions.  The application to discrete-basis
wave functions is shown at the end.
The total many-body wave function in the multi-channel GCM is given by, 
\begin{equation}
\Psi(x)=\sum_i\int dq\,f_i(q)\psi_{i,q}(x),
\end{equation}
where $i$ is a label for the states on a site and $q$ is a generator coordinate. 
The Hill-Wheeler equation for the weight function $f_i(q)$ reads,
\begin{equation}
    \int dq'\,\sum_{i'}H_{ii'}(q,q')f_{i'}(q')
=E\int dq'\,\sum_{i'}N_{ii'}(q,q')f_{i'}(q'),
\label{eq-app:HW}
\end{equation}
with 
$H_{ii'}(q,q')=\langle\psi_{i,q}|H|\psi_{i',q'}\rangle$ and 
$N_{ii'}(q,q')=\langle\psi_{i,q}|\psi_{i',q'}\rangle$.  

Here we assume translational symmetry requiring the Hamiltonian and the overlap 
matrices to depend only on the separate of grid points  $q-q'$. That is, $H_{ii'}(q,q')=H_{ii'}(0,q'-q)$ and 
$N_{ii'}(q,q')=N_{ii'}(0,q'-q)$. The solution of the Hill-Wheeler equation, (\ref{eq-app:HW}), 
is then given by, 
\begin{equation}
    f_i(q)=f_ie^{ikq}
\end{equation}
where the energy $E=E(k)$ and the coefficient $f_i$ are determined by solving the 
generalized eigenvalue problem, 
\begin{equation}
    \sum_{i'}\tilde{h}_{ii'}f_{i'}=E(k)\sum_{i'}\tilde{n}_{ii'}f_{i'},
    \label{eq-app:HW2}
\end{equation}
where
\begin{eqnarray}
\tilde{h}_{ii'}
&=&\int dq\,H_{ii'}(0,q)e^{ikq}  \\
\tilde{n}_{ii'}
&=&\int dq\,N_{ii'}(0,q)e^{ikq}. 
\end{eqnarray}
The formula corresponding to Eq. (\ref{eq-app:HW2}) in the discrete-basis
formalism is the same with the definitions 
\begin{eqnarray}
\tilde{h}_{ii'}
&=&\sum_n  H_{ii'}(0,n \Delta x) e^{i n k \Delta x}  
\\
\tilde{n}_{ii'}
&=&\sum_n  N_{ii'}(0,n \Delta q)e^{in k \Delta x}.  
\end{eqnarray}

For the Hamiltonian discussed in the text, $N_s =2$ states are on a site
and Eq. (\ref{eq-app:HW2}) is the generalized eigenvalue equation with
matrices of dimension 2.  It can be solved analytically in terms of 
the elements of the two matrices, but the resulting quadratic equation is
not very informative.  We note that for discrete-basis Hamiltonian with $N_s=1$ the
Eq. (\ref{eq-app:HW2}) reduces to the simple form presented in previous 
publications.

\end{document}